\documentclass[aps,twocolumn,prl,showpacs,color,psfig,epsf]{revtex4}
\usepackage{amsmath}
\usepackage{color}
\usepackage{amsfonts}
\usepackage{epsf}
\usepackage{graphicx}
\usepackage{enumerate}
\usepackage{ulem}
\definecolor{red}{rgb}{1,0,0}
\definecolor{blue}{rgb}{0,0,1}

\begin{document}


\title{Pattern formation in polymerising actin flocks: spirals, spots and waves without nonlinear chemistry}
\author{T. Le Goff\footnote{thomas.le-goff@ed.ac.uk}, B. Liebchen\footnote{bliebche@staffmail.ed.ac.uk}, D. Marenduzzo\footnote{dmarendu@ph.ed.ac.uk}}
\affiliation{SUPA, School of Physics and Astronomy, University of 
Edinburgh, Peter Guthrie Tait Road, Edinburgh, EH9 3FD, UK}

\begin{abstract}
We propose a model solely based on actin treadmilling and polymerisation which describes many characteristic states of actin wave formation: spots, spirals and travelling waves. 
In our model, as in experiments on cell recovering motility following actin depolymerisation, we choose an isotropic low density initial condition; polymerisation of actin filaments then raises the density towards the Onsager threshold where they align. We show that this alignment, in turn, destabilizes the isotropic phase and generically induces transient actin spots or spirals as part of the dynamical pathway towards a polarized phase which can either be uniform or consist of a series of actin-wave trains (flocks). 
Our results uncover a universal route to actin wave formation in the absence of any system specific nonlinear biochemistry, and it may help understand the mechanism underlying the observation of actin spots and waves {\it in vivo}. They also suggest a minimal setup to design similar patterns {\it in vitro}.
\end{abstract}

\maketitle


Actin networks are highly dynamic subcellular structures which constitute a key component of the cytoskeleton of eukaryotic cells~\cite{alberts}. 
These cells can be viewed as crosslinked gels made up from actin filaments, i.e. semi-flexible protein polymers with persistence and contour lengths both typically in the $1-10$ $\mu$m range. 
Actin filaments are active polymers which function far from thermodynamic equilibrium, as they constantly turn over their components, actin monomers, through polymerisation and depolymerisation~\cite{alberts,bray}. Under physiological conditions, the actin cytoskeleton forms a cortex just below the cell membrane, and it exploits polymerisation to power cellular motility~\cite{frey}, e.g. when a cell crawls on a substrate. 

Actin filaments and networks self-organize into a variety of mesmerising patterns~\cite{bausch,gerisch,brettschneider}. {\it In vitro}, experiments have reported the formation of lanes, waves and spirals in systems where actin fibers of constant length walk on a carpet of immobilised molecular motors~\cite{bausch}. {\it In vivo}, the actin network of a cell is normally localized within a $\mu$m-wide cortex trailing just behind the advancing membrane of a moving cell. However, under particular conditions, actin fibers reorganize within the cell, and create different patterns, such as travelling or scroll waves~\cite{gerisch,brettschneider,allard,actinwaveexpt1,actinwaveexpt2}. 

In some cases, the mechanism through which actin waves arise is relatively well understood, and is given by a network of biochemical regulatory reactions involving actin-associated proteins~\cite{Khamviwath}, which can be effectively modeled as an activator-inhibitor dynamical system. Such models, based on nonlinear biochemistry, successfully explain cases where actin waves are associated with the activation of the SCAR-WAVE complex~\cite{SCAR}, and they are linked to chemotaxis~\cite{chemotaxis}. However, there are other examples where waves depend on only a small number of components. Most relevant to our work are the waves observed in {\it Dictyostelium} cells recovering from treatment with latrunculin, which causes mass depolymerisation of actin fibers~\cite{gerisch,brettschneider}.  When latrunculin is taken away, actin fibers repolymerise from monomers in the cytosol, and after this cells recover motility: they do so by undergoing a surprisingly complex pattern formation cascade. First, actin assembles into transient spots, which then evolve into waves; spiral patterns are also observed in some cases. A set of experiments knocking out several actin-associated proteins clarified that the dynamics leading to waves is not dependent, among others, on the SCAR-WAVE complex, or on contractile myosin motors~\cite{brettschneider}.

The waves observed in Ref.~\cite{gerisch} have to date been addressed by a number of models in the literature~\cite{whitelam,pre,carlsson,kruse,beta}. All these works lead to wave formation, and all include some nonlinear dynamics, such as the Fitz-Nagumo model~\cite{whitelam}, or other activator-inhibitor models~\cite{pre}. This choice is often motivated by the observation that some actin-associated proteins {\it are} found in waves -- most notably, coronin, which localizes at the rear of a wave, and myosin I, which lies at the front~\cite{brettschneider}. While these are all perfectly plausible models, they either rely on the existence of a delay, or on (cubic) nonlinear reaction terms which are generally quite system specific.

Here we suggest an alternative model for wave formation, which does not require {\it any} nonlinear biochemistry, and solely depends on three simple and generic ingredients: actin polymerisation, steric repulsion between actin fibres, and treadmilling (i.e., the effective motion of actin fibres which grow at one end and shrink at the other one~\cite{alberts,bray}).
Since all three ingredients occur in a wide class of systems featuring actin waves, our findings suggest that spots and waves
could hinge on a universal mechanism and do not, as the current literature suggests, require system specific nonlinear chemistry. This key finding should be of particular relevance for the current understanding of waves in {\it Dictyostelium} {\it in vivo}; they also suggest how to set up experiments {\it in vitro} to generate similar patterns.

In our model actin filaments ``flock''~\cite{TonerPRL}: they align when dense enough, due to excluded volume interactions (like rigid rods in the Onsager theory for nematic liquid crystals~\cite{onsager}), and they move due to treadmilling, leading to actin waves.
At the low initial fiber densities typical of the early stages of experiments in {\it Dictyostelium}, however, alignment interactions are ineffective. As a preliminary step to wave formation, polymerisation increases the fibre density. Here, we unveil that spot formation, which is frequently observed in experiments prior to waves \cite{gerisch}, does not require complex reaction-based instabilities but occurs {\emph generically} as part of the dynamical pathway from the isotropic to the flocking phase. Here polymerisation shapes the morphology of the emerging waves, and allows controlling their lengthscale. 

To specify our qualitative arguments, we now propose a dynamical model to study pattern formation in a system of polymerising actin fibers, where we follow both the density of F-actin filaments, $\rho$, and their average polarisation  (i.e., the sum of orientation unit vectors per unit volume), ${\bf P}$. The equations of motion defining our model read as follows:
\begin{eqnarray}\label{eqofmot}
  \partial_t\rho & = &   - v_0{\bf \nabla}.(\rho{\bf P})+D_{\rho}{\bf \nabla}^2\rho+\alpha\rho\left(1-\frac{\rho}{\rho_0}\right) 
 \label{eom1}\\
  \partial_t{\bf P} & = & \gamma\left(\frac{\rho}{\rho_c}-1\right){\bf P}+K{\bf \nabla}^2{\bf P}-\gamma_2P^2{\bf P} \label{eom2}. 
\end{eqnarray}
Here $D_{\rho}$ is the diffusion coefficient for F-actin, $K$ is an effective elastic constant, while $v_0$ and $\alpha$ denote the treadmilling speed and the polymerisation rate respectively. Further, $\gamma$ measures how fast F-actin filaments change their direction, the term in $\gamma_2$ ensures saturation of the polarisation, whereas $\rho_c$ and $\rho_0$ indicate respectively the critical density above which nematic order sets in, and the target polymerisation density (i.e., the density of F-actin which would be reached due to polymerisation in a well-stirred system in the absence of spatial effects). For $\alpha=0$, Eqs.~(\ref{eom1},\ref{eom2}) are related to the models of Refs.~\cite{cristina,Rao,tonertu,caussin}, although even in that limit our emphasis here is on the dynamical pathway the system follows, rather than on steady state behaviour.
 
\begin{figure}
\centering
\def\svgwidth{0.9\columnwidth}
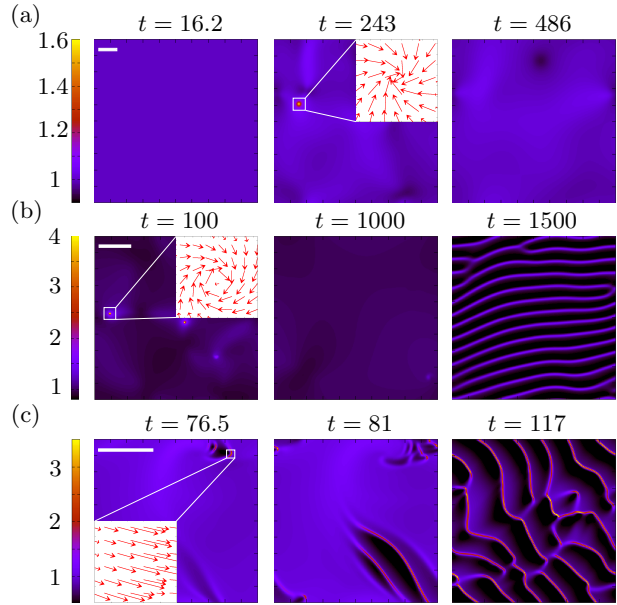
\caption{Representative snapshots for actin pattern formation, time increasing from left to right. (a) $\Gamma=1$: an actin spot forms and then disappears. (b) $\Gamma=4.3$: spiralling spots form early on; they then decay and are replaced by a regular wave train. (c) $\Gamma=10$: spots are polarised, and the final actin waves are irregular. Other parameters: $r=1.1$, $\mathcal{D}=5$ and $\Gamma_2=0.075$. The scale bar is 50.} 
\label{fig1}
\end{figure}

It is also useful to recast Eqs.~(\ref{eom1},\ref{eom2}) in terms of dimensionless variables, as follows,
\begin{eqnarray}\label{dimensionlesseqofmot}
  \partial_t\rho=-{\bf \nabla}.(\rho{\bf P})+{\bf \nabla}^2\rho+\rho\left(1-\rho\right)\label{eomdimless1}\\ 
  \partial_t{\bf P}=\Gamma\left(r\rho-1\right){\bf P}+\mathcal{D}{\bf \nabla}^2{\bf P}-\Gamma_2P^2{\bf P},\label{eomdimless2}
\end{eqnarray}
where we have defined $\Gamma=\gamma/\alpha$, $r=\rho_0/\rho_c$, $\mathcal{D}=K/D_{\rho}$, $\Gamma_2=\gamma_2D_{\rho}/v_0^2$, and we have further redefined $t\rightarrow\alpha t$, $x\rightarrow(\alpha/D_{\rho})^{1/2}x$, $\rho\rightarrow\rho/\rho_0$ and ${\bf P}\rightarrow(v_0/\sqrt{D_{\rho}\alpha}){\bf P}$, so as to have dimensionless time, space, density and polarisation. Eqs.~(\ref{eomdimless1},\ref{eomdimless2}) also clarify that the dynamics of our model depends on four dimensionless parameters -- while we have varied all of these, we have found that $\Gamma$, which is the ratio between alignment and polymerisation rate, is our key control parameter (provided that $r>1$). 
To provide an overview over the possible patterns in this system, we vary this parameter in the following while keeping other parameters at values given in the caption of Fig.~\ref{fig1}. It is useful to estimate the orders of magnitude of parameter values which are relevant to experiments. {\it In vivo} or in the lab, actin may polymerise at a rate $\alpha \sim 1-100$ s$^{-1}$~\cite{bray,xingbo}, while $\gamma$ may be estimated as the rotational diffusion of an intracellular F-actin filament of typical geometry $\sim 1 \mu$m $\times \sim 5$ nm, which is $\sim 10$ s$^{-1}$~\cite{xingbo}. For this geometry, the Onsager threshold of actin fibers can be estimated as $0.5\%$ in volume fraction -- the inverse of their aspect ratio -- whereas the fiber density in a cell is up to $\sim 10$ g/l~\cite{Factindensity}, or $\sim 1\%$ in volume fraction. As a result, an experimentally relevant range of parameters is $\Gamma\sim 0.1-10$, and $r>1$.  

\begin{figure}
\centering
\def\svgwidth{0.9\columnwidth}
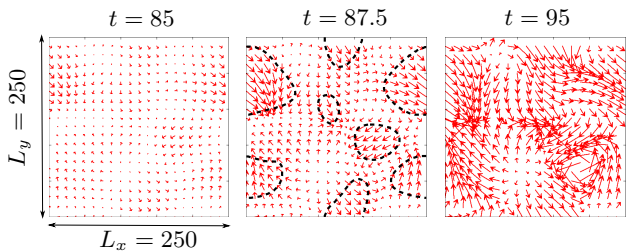
\caption{Snapshots of the evolution of the {\bf P}-field during the formation of spots for $\Gamma=4.3$, $r=1.1$, $\mathcal{D}=5$ and $\Gamma_2=0.075$, shown in Fig. 1b.}
\end{figure}

We have solved Eqs.~(\ref{eomdimless1},\ref{eomdimless2}) for different values of $\Gamma$ on a square lattice of size $Lx\times Ly$ using finite difference methods, periodic boundary conditions and a uniform initial state $\{\rho, {\bf p}\}=(0,{\bf 0})$ plus some small fluctuations. 
For identical polymerisation and alignment rate ($\Gamma=1$), we initially observe a uniform density growth followed, after a certain lag time, by the formation of one or several spots growing out of the uniform phase (Fig.~\ref{fig1}a and video 1 in SM). These spots have a spiral-like orientation of the actin fibres (Fig. 1a, inset). Remarkably, they are not stable, but decay after a lifetime of about 200 polymerisation cycles back to the uniform state. 
If fibres align faster than new ones are polymerised ($\Gamma=4.3$), we again observe transient spot formation. Intriguingly, however, here we do not end up with a uniform phase but observe the emergence of travelling actin waves. These waves self-arrange into a pattern with a well-defined length scale (Fig. ~\ref{fig1}b and video 2 in SM).  
Further enhancing the alignment rate ($\Gamma=10$) again leads to the formation of spots. Here, however, the spots are less pronounced and start to spiral and move while growing~\cite{xingbo}; they continuously transforms into travelling waves (Fig.~\ref{fig1}c and video 3 in SM). Further enhancing $\Gamma$ directly leads to waves without a preceding spot stage. Therefore, strikingly, our simple and generic model accounts for the sequence of actin patterns, from spots to waves, observed experimentally~\cite{gerisch,brettschneider}. We now want to understand why spots and then waves emerge. 

\begin{figure}
\centering
\def\svgwidth{0.9\columnwidth}
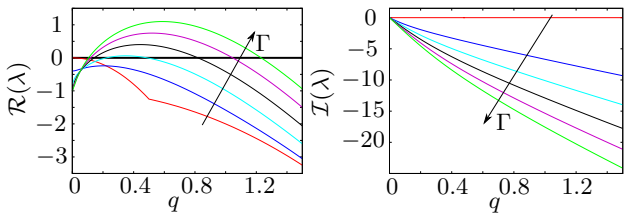
\caption{Real and imaginary parts of dispersion relation of small fluctuations around the uniform phase, for $\Gamma$ from 0 to 10 with $r=1.1$, $\mathcal{D}=5$ and $\Gamma_2=0.075$.}
\label{fig2}
\end{figure}

To this end, we now perform a linear stability analysis of our equations of motion (providing results here in physical units).
We note that the present system has three uniform solutions. These are: (i) $(\rho,{\bf p})=(0,{\bf 0})$ (which we chose as our initial state, following {\it in-vivo} experiments), (ii) $(\rho, {\bf p})=(\rho_0,{\bf 0})$, (iii) $(\rho,{\bf p})=(\rho_0,\sqrt{(\gamma/\gamma_2) (\rho_0/\rho_c-1)}{\bf e})$, where ${\bf e}$ represents a unit vector pointing in a spontaneously chosen direction set by the initial conditions. All solutions correspond to uniform phases: the first two are unpolarized, the third is polarized, hence travelling (flocking). First, we explore the stability of our initial low density state. The dominant branch of the dispersion relation of fluctuations around this phase reads $\lambda=\alpha -D_\rho {\bf q}^2$: therefore our initial state is generally unstable against polymerisation, simply leading to a density growth in the whole system if $\alpha>0$ (with no effect on the polarization field, as the eigenmode of the unstable mode is orthogonal to ${\bf p}$). This density growth proceeds until we have $\rho=\rho_c$; i.e. polymerisation generally transfers the system from phase (i) to phase (ii). 

Conversely to phase (i), for $\rho_0>\rho_c$, alignment interactions become effective in phase (ii), i.e. they dominate over rotational diffusion -- see Eq.~\ref{eom2}. 
This can be seen from the dominant branch of the dispersion relation, $\lambda = \gamma (\rho_0/\rho_c-1) -K{\bf q}^2$ (see SM), of fluctuations in this phase, which yields a stationary long wavelength instability.
Notably, alignment interactions are strong enough here to generate an instability of the uniform unpolarized phase but too weak to generate waves (which would require an oscillatory instability).
Following this instability, the dynamical pathway of our system is subtle and can be described as follows. 
Actin fibres align locally, leading to polarized domains, with the polarization field of each domain pointing in a spontaneously chosen direction. Due to treadmilling each of these domains moves, but soon 'collides' with other domains of aligned fibres, resulting in a defect in the ${\bf p}$ field with ingoing fibre-density flux from all directions (see Fig. 2), which in turn generates a spot in the density field (Fig.~\ref{fig1}a,b). 
This scenario is a natural and generic consequence of the instability of phase (ii) and therefore part of the dynamic pathway followed by our system, when initialised in phase (i).
We determined the length scale of the spots, $l$, by a combination of linear stability analysis (see SM) and systematic parameter sweeps, and found that $l \sim \sqrt{K/(\gamma(\rho_0/\rho_c-1))} (\gamma_2 D_\rho/v_0^2)^{1/4}$. Hence, the typical spot size increases with diffusion but decreases with self-propulsion velocity. This scaling is intuitive, since fibres treadmill from all directions towards the spot center thereby competing with diffusion (a similar scaling, albeit leading to a distinct functional form for $l$, determined the size of aster size in~\cite{Rao}).
Remarkably, we found that for $\gamma_2 D_\rho/v_0^2>1$ the spot size converges to a healing length $l \sim \sqrt{K/(\gamma(\rho_0/\rho_c-1))}$ representing the distance needed for the polarization field to recover from a local orientational perturbation (defect).
We note that, in the absence of polymerisation ($\alpha=0$), our asters satisfy the steady state condition of Eq.~\ref{eom1}  ($\dot \rho=0$) yielding a solution ${\bf p}\propto \nabla \rho/\rho$ where fibre treadmill up the density gradient thereby permanently balancing diffusive fibre losses. Importantly, however, the local density in the spot exceeds $\rho_0$, leading for $\alpha >0$ to depolymerisation. This in turn initiates Fisher waves travelling from the spot in all directions. We expect these Fisher wave fronts to move with a characteristic velocity of $v=\sqrt{2D_\rho\alpha}$; these waves combine with the alignment interactions to decrease spot size and take the system back towards a uniform phase. 
This scenario describes the transition from a uniform phase to spot and back to uniformity as observed in Fig.~\ref{fig1}a. But why does the described scenario not repeat to initiate new spots? The answer is, that the new uniform phase is now polarized and given by (iii) rather than by (ii). Hence, the spots in Fig.~\ref{fig1}a (video 1 in SM) are a generic transient pattern formed as actin fibers polymerise starting from a low density phase.

Having followed this pathway from phase (i) to phase (iii), we now want to know how waves emerge. Let us therefore explore the stability of phase (iii) by calculating the dispersion relation of fluctuations around this phase (Fig.~\ref{fig2}, see also SM). 
Remarkably, $\mathcal{R}(\lambda)$ is always negative at small $q$ but becomes positive at finite $q$ if $\Gamma$ is sufficiently large (Fig.~\ref{fig2}). Since also the imaginary part of the dispersion relation is finite, we have an {\it oscillatory} {\it short wavelength} instability and may therefore expect travelling waves for sufficiently strong alignment interactions ($\Gamma$): this explains our previous observation of travelling waves in Figs.~\ref{fig1}b and Fig.~\ref{fig1}c. 
We find that the velocity of our waves is given by $v\sim v_0 \sqrt{(\gamma/\gamma_2) (\rho_0/\rho_c -1)}=v_0 P_0$ -- this is true if $r=(\rho_0/\rho_c -1)>1$, otherwise $v\sim v_0\sqrt{\gamma/\gamma_2}$. Therefore the wave speed is proportional to the treadmilling velocity of individual fibres, weighted by an alignment factor measuring the average fraction of aligned filaments. The distance between adjacent wave peaks can be estimated by numerical evaluation of the dispersion relation of our linear stability analysis revealing a fastest growing mode at length scale 
$l\propto l_1^{3/2}l_2^{-1/2}$ with $l_1\sim \sqrt{K/\gamma}$ and $l_2\sim \sqrt{K D_{\rho}/(v_0^2P_0^2)}$ if $K>D_{\rho}$, or $l_2\sim \sqrt{K^3 /(D_{\rho}v_0^2P_0^2)}$ if $K<D_{\rho}$ (see SM, note we have dropped for simplicity an extra non-dimensional dependence on $r$). 
Our simulations confirm that the wave length is typically close to this value, at least deep into the wave forming regime. The width of our wave peaks follows, approximately, a variant of our healing length $l\sim \sqrt{K/\gamma}$ for $K>D_\rho$ and $l \sim \sqrt{D_\rho/\gamma}$ if $D_\rho>K$ (see SM for more details) 
-- in this context this is the length scale over which diffusion neutralizes polar ordering within a wave peak. 

\begin{figure}
\centering
\def\svgwidth{0.9\columnwidth}
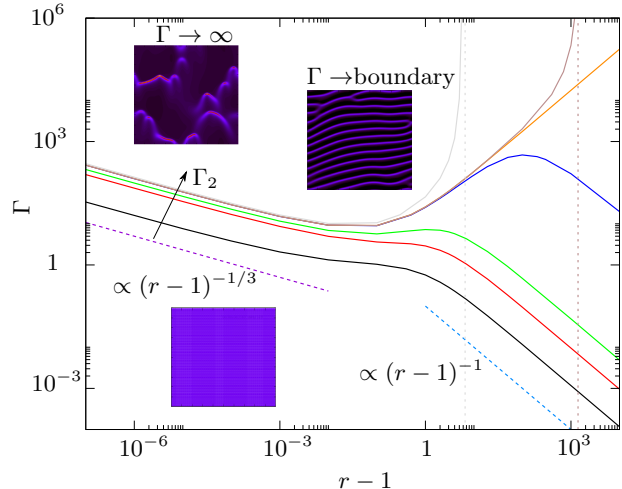
\caption{Phase diagram in the $(\Gamma,r-1)$ plane. Curves correspond to $\mathcal{D}=5$ and selected values of $\Gamma_2$. From top to bottom, these are: $0.00075$, $0.075$, $0.1775$, $0.3704$, $3/8$, $0.3754$, $0.4688$.}
\label{fig3}
\end{figure}

The possible scenarios can be summarised in a phase diagram (Fig.~\ref{fig3}). For small $\Gamma$ and $r-1$, i.e. when alignment interactions are weak and the saturation fibre density is close to the Onsager threshold, the evolution features a uniform increase of our low density initial state, i.e. a transition from phase (i) via (ii); this phase then morphs into a set of asters and spirals, which leave way eventually to a phase (iii) which is asymptotically stable. 
Instead, when we cross the transition line, along the black arrow in Fig.~\ref{fig3}, we always find travelling actin waves at long timescales. 
Deep in the wave phase (large $\Gamma$), we find waves emerging directly within the uniform phase: these waves are also irregular and peaks are far from each other. Closer to the transition line we typically find spots appear before waves emerge; waves are also more regular and the separation between peaks can be decreased, for instance, by increasing polymerisation (hence decreasing $\Gamma$). 
The different dynamics occur since close to the transition line, waves emerge slowly and do not impede spot formation on the pathway from phase (ii) to phase (iii). 
The length scale selection may be linked to the fact that the longest wavelength in the instability band (where the real part of the dispersion relation is positive) depends on $\alpha$.
For $\Gamma_2>3/8$ (orange curve) there is an additional transition line in our phase diagram (dashed lines for grey and brown line), 
representing a parameter domain of large actin fiber density where waves are impossible even for very strong alignment interactions. Physically, this means that wave formation is only 
possible in our system if self-propulsion is fast enough; the critical speed is given by a combination of fiber diffusion and alignment saturation.   

Finally, we would like to highlight here the role of polymerisation in our model. Besides guiding the system through successive instabilities as the overall density increases, actin polymerisation plays other major roles in the system. First, when its rate $\alpha$ is large enough, it can suppress pattern formation altogether. Second, close to the instability threshold, 
the polymerisation rate controls the width of and separation between wave peaks. Finally, at least within the range which we have explored, polymerisation is 
required to create transient spots.

To conclude, we have shown that an ensemble of polymerising and treadmilling actin filaments forms a cascade of patterns encompassing spots, spirals and waves, which resemble the typical phenomenology found in experiments. Specifically, when {\it Dictyostelium} cells recover from actin depolymerisation, they reassemble their actin cytoskeleton by creating spots which later on transition to waves~\cite{brettschneider,whitelam}. 
Remarkably, and at variance with previous work, our model can recreate this sequence of patterns without the need to assume any underlying nonlinear biochemistry leading to delay, or oscillatory or activator-inhibitor behaviour. 
Instead, starting from a low density initial phase, we suggest that polymerisation increases the overall density of actin until locally oriented domains of moving actin flocks appear. These domains travel along randomly selected directions, and collide with each other to form spirals or larger spots where the filament directions are arranged in an aster shape. 
Hence, our work demonstrates that spots occur automatically en route from the typical low density initial phase towards the flocking state featuring waves, thereby challenging previous and more complicated mechanisms describing the phenomenology of typical {\it in vivo} actin wave experiments. Besides this, our results might also be useful for designing and understanding minimal {\it in vitro} systems mimicking the actin dynamics observed {\it in vivo}.

We thank EPSRC (grant EP/K007404/1) for support. BL gratefully acknowledges funding by a Marie Sk{\l}odowska Curie Intra European Fellowship (G.A. no 654908) within Horizon 2020.

\newpage


\section{Supplementary Material}

Here, we perform a detailed linear stability analysis of our model to derive the expressions for the stability criteria and length scale which we 
used in the discussion of spot and wave formation in the main text.
To keep our calculations comprehensive, we consider the equations of motion in dimensionless form (see main text):
\begin{equation}
\left\{
\begin{array}{l}
  \partial_t{\bf P}=\Gamma\left(r\rho-1\right){\bf P}+\mathcal{D}{\bf \nabla}^2{\bf P}-\Gamma_2P^2{\bf P}, \\
  \partial_t\rho=-{\bf \nabla}\cdot(\rho{\bf P})+{\bf \nabla}^2\rho+\rho\left(1-\rho\right), 
\end{array}
\right.
\label{equasystemdimensionless}
\end{equation}
where $\Gamma=\gamma/\alpha$, $r=\rho_0/\rho_c$, $\mathcal{D}=K/D_{\rho}$ and $\Gamma_2=\gamma_2D_{\rho}/v_0^2$. 

These equations allow us to identify three different stationary uniform solutions for the density and polarization field: 
\begin{enumerate}[(i)]
\item ${\bf P}={\bf 0}$ and $\rho=0$,
\item ${\bf P}={\bf 0}$ and $\rho=1$,
\item ${\bf P}={\bf P_0}=[(\Gamma/\Gamma_2)(r-1)]^{1/2}{\bf e}$ and $\rho=1$,
\end{enumerate}
Solutions (i) and (ii) represent uniform isotropic phases with zero and finite density, respectively. 
In contrast, (iii) is a uniform polarized phase with spontaneously chosen polarization direction ${\bf e}$. This phase only exists if $r\geq1$, i.e. if the target polymerization density is larger than the critical density at which alignment interactions dominate over rotational diffusion of 
the actin filaments. 
We will now test the stability of each of the phases (i)--(iii) against small fluctuations. Therefore we linearize Eqs.~(\ref{equasystemdimensionless}) around solutions (i)--(iii) respectively, 
and solve the resulting equations in Fourier space (or alternatively by plugging a plane wave ansatz $({\bf P},\rho)=({\bf P_s},\rho_s)+(A_{{\bf P}},A_{\rho})\exp(\lambda t+i{\bf q}.{\bf r})$ 
with small $A_{{\bf P}}$ and $A_{\rho}$ into the linearized equations).

\subsection{Uniform growth}
Following typical experiments, in our simulations we initialized our system in phase (i) with some additional fluctuations in the particle density field (see main text). 
To follow the dynamics of our system, we investigate the stability of phase (i) first. 
Solving the linearized version of Eqs.~(\ref{equasystemdimensionless}) around phase (i) in Fourier space, quickly leads to the following condition for the existence of plane wave solutions:
$$
\begin{vmatrix}
\lambda+\mathcal{D}q^2+\Gamma& 0& 0\\
0& \lambda+\mathcal{D}q^2+\Gamma& 0\\
0& 0& \lambda+q^2-1
\end{vmatrix}=0.
$$
From this condition we quickly determine the dispersion relation $\lambda(q)$ for plane wave fluctuations, whose largest branch is:
\begin{equation}
  \lambda=-q^2+1
\end{equation}
Translation this result back to physical units leads to $\lambda=-D_{\rho}q^2+\alpha$, which shows that polymerization creates a long wavelength instability of phase (i).
Notably, the eigenmode corresponding to this unstable growth is orthogonal to ${\bf P}$ leading to a simple growth of the 
actin filament concentration without aligning filaments, tranferring our system from phase (i) to (ii).

\subsection{Spots formation}
Now analysing the linear stability of phase (ii) we find the condition
$$
\begin{vmatrix}
\lambda+\mathcal{D}q^2-\Gamma\left(r-1\right)& 0& 0\\
0& \lambda+\mathcal{D}q^2-\Gamma\left(r-1\right)& 0\\
iq_x& iq_y& \lambda+q^2+1
\end{vmatrix}=0.
$$
yielding for the largest branch of the dispersion relation 
\begin{equation}
  \lambda=-\mathcal{D}q^2+\Gamma\left(r-1\right)
\end{equation}
Translating this expression back to physical variables, we find
$ \lambda=-Kq^2+\gamma\left(\frac{\rho_0}{\rho_c}-1\right)$ featuring another long wavelength instability 
if $\rho_0>\rho_c$.
Remarkably, this instability is now parallel to ${\bf P}$ meaning that once the system has reached phase (ii) alignment interactions become effective but do not affect the density field in the linear regime. 
We can observe a corresponding alignment of actin filaments in 
Fig.2 of the main text generically leading to the formation of defects in ${\bf p}$ which results in formation of spots in ${\bf \rho}$ which is a purely nonlinear effect
and part of the dynamical pathway of our system from phase (ii) to (iii) (rather than a distinct 'spot' phase).

\subsection{Generation of waves}
To analyse the linear stabilty of phase (iii) we choose a coordinate system where the direction of polarization ${\bf e}$ is parallel to the $x$-axis, i.e. where 
${\bf P}_0=P_0{\bf e_x}=[(\Gamma/\Gamma_2)(r-1)]^{1/2}{\bf e_x}$. Here, linear stability analysis leads to the condition

$$
\begin{vmatrix}
\lambda+\mathcal{D}q^2+2\Gamma\left(r-1\right)& 0& -\Gamma rP_0\\
0& \lambda+\mathcal{D}q^2& 0\\
iq_x& iq_y& \lambda+q^2+1+iP_0q_x
\end{vmatrix}=0
$$
yielding the following implicit equation for the dispersion relation of plane wave fluctuations
\begin{equation}
\begin{split}
&\left\{\left[\lambda+\mathcal{D}q^2+2\Gamma\left(r-1\right)\right]\left(\lambda+q^2+1+iP_0q_x\right)+irq_x\Gamma P_0\right\}\\
&\times \left\{\lambda+\mathcal{D}q^2\right\}=0. \label{charp}
\end{split}
\end{equation}
Ignoring the last term which generates only a negative solution for $\lambda$ in which we are not interested, Eq.~(\ref{charp}) can be rewritten as
\begin{equation}
\begin{split}
&\lambda^2+\left[\left(\mathcal{D}+1\right)q^2+iP_0q_x+2\Gamma\left(r-1\right)+1\right]\lambda+\mathcal{D}q^4\\
&+i\mathcal{D}P_0q_xq^2+\left[\mathcal{D}+2\Gamma\left(r-1\right)\right]q^2+i\Gamma\left(3r-2\right)P_0q_x\\
&+2\Gamma\left(r-1\right)=0.
\end{split}
\end{equation}
This equation leads to the following dispersion relation:
\begin{equation}
\lambda_{\pm}=-\frac{1}{2}\left[\left(\mathcal{D}+1\right)q^2+iP_0q_x+2\Gamma\left(r-1\right)+1\right]\pm f
\end{equation}
Here $f$ is a complex number. Since the product of roots of a polynomial function of the form $P_{ol}(x)=\sum_{i=0}^Na_ix^i$ is equal to $\frac{(-1)^Na_0}{a_N}$, here leading to $\lambda_+ \lambda_-=a_0$, we can write
\begin{equation}
\begin{split}
f^2=&\frac{1}{4}\left[\left(\mathcal{D}+1\right)q^2+iP_0q_x+2\Gamma\left(r-1\right)+1\right]^2-\mathcal{D}q^4\\
&-i\mathcal{D}P_0q_xq^2-\left[\mathcal{D}+2\Gamma\left(r-1\right)\right]q^2-i\Gamma\left(3r-2\right)P_0q_x\\
&-2\Gamma\left(r-1\right).
\end{split}
\end{equation}
Real and imaginary parts of $f^2$ are then easily calculated
\begin{eqnarray}
  F=\mathcal{R}(f^2)&=&\left[\frac{(\mathcal{D}-1)}{2}q^2+\Gamma\left(r-1\right)-\frac{1}{2}\right]^2-\left(\frac{P_0}{2}q_x\right)^2,\nonumber\\
  G=\mathcal{I}(f^2)&=&-P_0q_x\left[\frac{(\mathcal{D}-1)}{2}q^2+\Gamma\left(2r-1\right)-\frac{1}{2}\right].\nonumber
\end{eqnarray}
We finally obtain
\begin{eqnarray}
f= & & \left[\frac{F}{2}+\frac{1}{2}\left(F^2+G^2\right)^{1/2}\right]^{1/2} \\ \nonumber & + & i\frac{G}{2\left[\frac{F}{2}+\frac{1}{2}\left(F^2+G^2\right)^{1/2}\right]^{1/2}}.
\end{eqnarray}
Note there are two solutions for $f$ but both solutions lead to the same growth rates which read
\begin{eqnarray}
  \lambda_+=-\frac{(\mathcal{D}+1)}{2}q^2-\frac{1}{2}-\Gamma\left(r-1\right)-i\frac{P_0q_x}{2}+f,\\
  \lambda_-=-\frac{(\mathcal{D}+1)}{2}q^2-\frac{1}{2}-\Gamma\left(r-1\right)-i\frac{P_0q_x}{2}-f,
\end{eqnarray}
In physical variables these dispersion relations read:
\begin{eqnarray}
  \lambda_+ =  & - & \frac{(K+D_{\rho})}{2}q^2-\frac{\alpha}{2}-\gamma\left(\frac{\rho_0}{\rho_c}-1\right) \\ \nonumber & - &i\frac{v_0P_0q_x}{2}+f,\\
  \lambda_-= & - &\frac{(K+D_{\rho})}{2}q^2-\frac{\alpha}{2}-\gamma\left(\frac{\rho_0}{\rho_c}-1\right) \\ \nonumber 
& - &i\frac{v_0P_0q_x}{2}-f.
\end{eqnarray}
We note, that the imaginary part of $\lambda$ is different from zero. Hence, phase (iii) is subject to an oscillatory instability allowing for moving patterns. 
In contrast to the instabilities of phases (i) and (ii) which generically occur for $\alpha>0$ and large densities $r>1$ phase (iii) can be stable even at large densities and for all values of $\alpha>0$. 
If $\Gamma$ is sufficiently small 
the real part of our dispersion relation is negative; then we only observe spot formation and our system ends up in phase (iii) without featuring waves. 

The fastest growing mode for the dispersion relation corresponding to the generation of waves can be found as the value of $q$ for which the real part of $\lambda$ is maximal. From a numerical analysis, we find that the corresponding wavelength $l$ obeys the scaling:
\begin{equation}
l\sim l_1^{3/2}l_2^{-1/2}
\end{equation}
where the two lengthscales $l_1$ and $l_2$ depend on parameters as follows, 
\begin{eqnarray}
l_1 & \sim & [K/(\gamma(\rho_0/\rho_c-1))]^{1/2} \, \, \, \, \, \, \, \, \, {\rm if} \, \rho_0/\rho_c-1>1,
\\  
l_1 & \sim & [K/(\gamma(\rho_0/\rho_c-1)^{5/9})]^{1/2} \, {\rm if} \, \rho_0/\rho_c-1<1, \nonumber
\\ 
l_2 & \sim & [KD_{\rho}/(v_0^2P_0^2)]^{1/2} \qquad \, \, \, {\rm if} \, K>D_{\rho}, \\
l_2 & \sim & [K^3/(D_{\rho}v_0^2P_0^2)]^{1/2} \qquad {\rm if} \, K<D_{\rho}. \nonumber
\end{eqnarray}






\begin{thebibliography}{99}
\bibitem{alberts} B. Alberts, A. Johnson, J. Lewis, M. Raff, K.~K. Roberts and P. Walter, {\it Molecular biology of the cell} (Garland Science, New York, 2002).
\bibitem{bray} D. Bray, {\it Cell movements} (Garland Science, New York, 2000).
\bibitem{frey} A. Gholami, M. Falcke and E. Frey, {\it New J. Phys.} {\bf 10}, 033022 (2008).
\bibitem{bausch} V. Schaller, C. Weber, C. Semmrich, E. Frey and A.~R. Bausch, {\it Nature} {\bf 467}, 73 (2010); J.~F. Joanny and S. Ramaswamy, {\it Nature} {\bf 467}, 33 (2010).
\bibitem{gerisch} G. Gerisch, T. Bretschneider, A. M\"uller-Taubenberger, E.
Simmeth, M. Ecke, S. Diez and K. Anderson, {\it Biophys. J.} {\bf 87}, 3493 (2004).
\bibitem{brettschneider} T. Bretschneider, K. Anderson, M. Ecke, A.~M. M\"uller-
Taubenberger, B. Schroth-Diez, H.~C. Ishikawa-Ankerhold and
G. Gerisch, {\it Biophys. J.} {\bf 96}, 2888 (2009).
\bibitem{allard} J. Allard and A. Mogilner, {\it Curr. Opin. Cell Biol.} {\bf 25}, 107 (2013).
\bibitem{actinwaveexpt2} M.~G. Vicker, {\it Biophys. Chem.} {\bf 84}, 87 (2000).
\bibitem{actinwaveexpt1} M.~G. Vicker, {\it Exp. Cell Res.} {\bf 275}, 54 (2002).
\bibitem{Khamviwath} V. Kamviwath, J. Hu and H.~G. Othmer, {\it PLoS ONE} {\bf 8}, e64272 (2013).
\bibitem{SCAR} A. Y. Pollitt and R. H. Insall, {\it J. Cell Sci.} {\bf 122}, 2575 (2009). 
\bibitem{chemotaxis} O.~D. Weiner, W.~A. Marganski, L.~F. Wu, S.~J. Altschuler and M.~W. Kirschner, {\it PLoS Biol.} {\bf 5}, e221 (2007).
\bibitem{whitelam} S. Whitelam, T. Bretschneider and N.~J. Burroughs, {\it Phys. Rev. Lett.} {\bf 102}, 198103 (2009).
\bibitem{pre} V. Wasnik and R. Mukhopadhyay, {\it Phys. Rev. E} {\bf 90}, 052707 (2014).
\bibitem{carlsson} A.~E. Carlsson, {\it Phys. Rev. Lett.} {\bf 104}, 228102 (2010).
\bibitem{kruse} K. Doubrovinski and K. Kruse, {\it Europhys. Lett.} {\bf 83}, 18003 (2008).
\bibitem{beta} C. Beta, {\it PMC Biophys.} {\bf 3}, 12 (2010).
\bibitem{TonerPRL} J. Toner, {\it Phys. Rev. Lett.} {\bf 108}, 088102 (2012).
\bibitem{onsager} L. Onsager, {\it Ann. N. Y. Acad. Sci.} {\bf 51}, 62 (1949).
\bibitem{cristina} S. Mishra, A. Baskaran and M.~C. Marchetti, {\it Phys. Rev. E} {\bf 81}, 061916 (2010).
\bibitem{Rao} A. Chaudhuri, B. Bhattacharya, K. Gowrishankar, S. Mayor and M. Rao, {\it Proc. Natl. Acad. Sci. U.S.A.} {\bf 108}, 14825 (2011); K. Gowrishankar and M. Rao, {\it Soft Matter} {\bf 12}, 2040 (2016).
\bibitem{tonertu} J. Toner, Y.-h. Tu, and S. Ramaswamy, {\it Ann. Phys.} {\bf 318}, 170 (2005).
\bibitem{caussin} J.-B. Caussin, A. Solon, A. Peshkov, H. Chat\'e, T. Dauxois, J. Tailleur, V. Vitelli and D. Bartolo, {\it Phys. Rev. Lett.} {\bf 112}, 148102 (2014).
\bibitem{xingbo} X. Yang, D. Marenduzzo and M.~C. Marchetti, {\it Phys. Rev. E} {\bf 89}, 012711 (2014).
\bibitem{Factindensity} M. Bailly {\it et al.}, {\it J. Cell. Biol.} {\bf 145},  331 (1999).
\end{thebibliography}
\end{document}